\documentclass[12pt]{article}
\usepackage[T1]{fontenc}
\usepackage[utf8]{inputenc}
\usepackage{lmodern}
\usepackage{amsmath}
\usepackage{amssymb}
\usepackage{graphicx}
\usepackage{caption}
\usepackage{subcaption}
\usepackage{multirow}
\usepackage{float}

\title{The Interaction of Two Dyons in The Near Field Limit }
\author{Motahareh Kiamari$^{1}$, Sedigheh Deldar $^{1}$ \\
$^1$Department of Physics, University of Tehran,\\
\small P.O. Box 14395/547, Tehran 1439955961,
Iran.}
\date{}
\begin{document}

\maketitle

\begin{abstract}
We study the interaction of two dyons in the region of their cores where
they are non-linear and non-Abelian. We assume the superposition of two
dyons as a solution of the equation of motion. The terms due
to the non-linearity of the strength tensor are considered as the perturbation terms
which deforms the profile function of two individual dyons. As a result, the
profile function of dyons are obtained to be dependent on the polar angle and the 
spherical symmetry is lost.
\end{abstract}

\section{Introduction}
\label{introduction}

The structure of QCD vacuum is an interesting subject to study in particle physics. Calorons are among the candidates for this structure. Calorons are the periodic solutions of Euclidean Yang-Mills theory which are classified with the holonomy as an order parameter. KvBLL instanton \cite{Krann98}\cite{Lee98} are described by maximally non-trivial holonomy.

In a set of papers \cite{Diak2009}\cite{Diak2007}\cite{Diak2004}, authors showed that confinement-deconfinement transition can be described by the KvBLL instantons. They studied non-interacting ensemble of KvBLL instantons or calorons by considering the Polyakov loop as an order parameter and calculated the free energy of the quark-antiquark pair and the transition temperature. 

To study the interacting ensemble of calorons, the superposition of calorons should be studied. ADHM method \cite{ADHM} \cite{Atiyah} is a powerful method to construct the arbitrary self-dual solutions of Yang-Mills theory which results the construction of KvBLL instanton with dyons at finite temperature \cite{Krann98}\cite{Lee98}. This method gives an exact and stable multi-dyon and multi-caloron configurations. However, since the superposition of calorons is not the solution of self-dual Yang-Mills equation, there were some attempts \cite{Gerh2004} to improve this superposition by studying the caloron-Dirac string interaction and quasi-ADHM method. The former method applied to the dissociated caloron, which can be described with its constituents: dyons, and a Dirac string. Then the interaction of an object on the top of the Dirac string, namely a dyon of another caloron, with this string was calculated. As a result, the superposition of a dissociated caloron and an object on top of its Dirac string was the sum of unchanged caloron and a rotated object. This method can be applied to multi dissociated calorons which represents "an improved superposition of multi-caloron configuration" and can be compared with the quasi-ADHM method \cite{Gerh2004}. 

Another way to improve the superposition of calorons is perhaps considering the interaction of the dyons of different dissociated calorons. Dyons have non-linear and non-Abelian $SU(2)$ cores; while outside their cores they are Abelian $U(1)$ objects with a preferred direction in the color space which contain coulomb like electric and magnetic fields. Hence, one can assume Coulombic interactions for them outside their cores. In this way, one can compare the new superposition with the improvement of caloron-Dirac string interaction and quasi-ADHM method \cite{Gerh2004}.

To study the non-interacting ensemble of calorons, one should consider a dyonic gas. Therefore, one must add the classical dyon fields in a gauge which minimizes the interaction of dyons -- at least at infinity \cite{Diak2009} \cite{Mart95}. This makes the superposition of the dyons to be as close as possible to the solution of the Yang-Mills equation, since the sum ansatz of the dyons is no longer the solution of the Yang-Mills equations. The proper gauge is the "stringy gauge" where the temporal gauge fields of dyons are directed to the third direction of the color space at infinity \cite{Diak2009}. In this gauge, the interaction of dyons is zero at infinity and the sum of the dyon fields is the solution of the Yang-Mills equation for infinity. However, to study an interacting ensemble, one should forget gas approximation and do the calculations in other gauges including hedgehog gauge, where interaction is neither zero nor minimal even at infinity.

In this paper we do not use gas approximation and study the interaction of two dyons in the core area. In fact we study the changes of the gauge fields of a dyon in the presence of another dyon, where their cores overlap. We suppose the superposition of two dyons as the solution of the self-dual Yang-Mills equation and find the equation of motion for each individual dyons which is the perturbed version of the original equation. In fact, we suppose the presence of a dyon as a perturbation term that perturbs the self-dual equation of the first dyon. This perturbation changes the gauge field of the first dyon as a function of the angle between the radius vector and the vector which connects the centers of two dyons. As a result, the dyon has no longer spherical symmetry. 

The paper is organized as the following. In section \ref{sec:dyon}, dyons are introduced briefly. In section \ref{sec:interaction}, we calculate the profile functions of a dyon in the presence of another dyon and compare it with the ones of an individual dyon. We conclude the paper in section \ref{sec:concl}.
%------------------------------------------------------------------------------------------------------------------------------------------------------------------

\section{Dyon in SU(2) Yang-Mills theory}
\label{sec:dyon}
A dyon is originally an $SU(2)$ object with both electric and magnetic charges. It is a static and time-independent self-dual solution of the Yang-Mills theory,
\begin{equation}
F^{a}_{\mu \nu} = \tilde{F}^{a}_{\mu \nu} \longrightarrow  E^{a}_{i} = B^{a}_{i},
\label{self-duality}
\end{equation}
where $F$ is the strength tensor of the $SU(2)$ gauge group 
\begin{equation}
F^{a}_{\mu \nu} = \partial _{\mu}A^{a}_{\nu} - \partial _{\nu}A^{a}_{\mu} + \epsilon _{abc} A^{b}_{\mu} A^{c}_{\nu},
\label{F}
\end{equation}
and 
\begin{equation}
E^{a}_{i} = F^{a}_{i4},
\label{E}
\end{equation}
\begin{equation}
B^{a}_{i} = \frac{1}{2}\epsilon _{ijk} F^{a}_{jk},
\label{B}
\end{equation}
are the electric and magnetic fields, respectively. Therefore, the time-independent self-dual equation is,
\begin{equation}
  F^{a}_{i4} - \frac{1}{2}\epsilon _{ijk} F^{a}_{jk} = 0.
\label{sd-equation}
 \end{equation}
In $SU(2)$, there exist two dyons with different electric and magnetic charges \cite{Diak2009}. The first is called M with electric and magnetic charges (+,+) and in hedgehog gauge, or to be more precise the Rossi \cite{PRossi82} gauge, is defined by,
\begin{equation}
A^{a}_{4}(x) = n_{a} \frac{E(r)}{r}, \;\;\;\; E(r) = 1 - \nu r \coth \nu r,
\label{A4}
\end{equation}
\begin{equation}
A^{a}_{i} (x) = \epsilon_{aij} n_{j} \frac{1 - A(r)}{r}, \;\;\;\;  A(r) = \frac{\nu r}{\sinh \nu r},
\label{Ai}
\end{equation}
where $E(r)$ and $A(r)$ are called profile functions and $n_{i}$ is a unit vector in 3D space and $r$ is the distance from the center of the dyon which is located at the origin. $\nu \equiv \sqrt{A^{a}_{4}A^{a}_{4}}|_{\lvert \textbf{x}\rvert \rightarrow \infty}$ is called holonomy, and specifies the confinement and deconfinement phases. The second dyon called L with electric and magnetic charges (-,-) is obtained by the replacement $ \nu \rightarrow 2 \pi T - \nu $. Using equations (\ref{A4}) and (\ref{Ai}), the electric and magnetic fields can be obtained,
\begin{equation}
E^{a}_{i} = \left( \delta _{ai} - n_{a}n_{i} \right) \frac{A(r)}{r}\frac{E(r)}{r} + n_{a}n_{i} \frac{d}{dr} \left( \frac{E(r)}{r} \right),
\label{E-field}
\end{equation}
\begin{equation}
B^{a}_{i} = \left( \delta _{ai} - n_{a}n_{i} \right) \frac{1}{r} \frac{dA(r)}{dr} + n_{a}n_{i} \frac{A^{2}(r) - 1}{r^{2}}.
\label{B-field}
\end{equation}

A dyon is non-linear and non-Abelian in the region of its core of size $ \sim \frac{1}{\nu} $. Therefore, 
\begin{subequations}
\begin{align}
        E(r) = 1 - \nu r \coth \nu r \xrightarrow{r\rightarrow 0} -\frac{(\nu r)^{2}}{3},\\
        A(r) = \frac{\nu r}{\sinh \nu r} \xrightarrow{r\rightarrow 0} 1 - \frac{(\nu r)^{2}}{6},
        \label{bc0-A}
\end{align}
\end{subequations}
where $\lim _{r\rightarrow 0} E(r) = 0$ and $\lim _{r\rightarrow 0} A(r) = 1$ are the boundary conditions to avoid singularity. For the far field limit, outside the core, a dyon is Abelian along an arbitrary direction,
\begin{subequations}
\begin{align}
        E(r) = 1 - \nu r \coth \nu r \xrightarrow{r\rightarrow \infty} -\nu r + 1 - O(e^{-\nu r}),\\
        A(r) = \frac{\nu r}{\sinh \nu r} \xrightarrow{r\rightarrow \infty} O(e^{-\nu r}).
\end{align}
\end{subequations}
One can "gauge comb" it to the third color direction \cite{Diak2009},
\begin{equation}
A^{a}_{i}\xrightarrow{r\rightarrow \infty} 0, \;\;\; A^{a}_{4} \xrightarrow{r\rightarrow \infty} -\nu \frac{\sigma ^{3}}{2} 
\end{equation}
where $\sigma ^{3}$ is the third Pauli matrix. This gauge is called "stringy gauge". Holonomy is related to the confinement-deconfinement transition by the order parameter Polyakov loop,
\begin{equation}
P(\textbf{r})=\frac{1}{2}Tr\left(\exp\left(i\int_{0}^{1/T} dx_{4}A_{4}\left(x_{4},\textbf{r}\right) \right) \right),
\label{polyakovloop}  
\end{equation}
where $ P(\textbf{r})\rightarrow 0 $ for maximally non-trivial holonomy and $\nu = \pi T$ which corresponds to the confinement phase. $ P(\textbf{r})\rightarrow \pm 1 $ for trivial holonomy that corresponds to the deconfinement phase.
%-----------------------------------------------------------------------------------------------------------------------------------------------------------------

\section{Interaction between two dyon cores}
\label{sec:interaction}
In this paper we study the interaction between two dyons in the region of their cores, where dyons are non-linear and non-Abelian. Outside this region, dyons are Abelian with a preferred direction in the color space that can be chosen to be the third direction.
 \begin{figure}
 \captionsetup{font=footnotesize}
  \begin{center}
    \includegraphics[width=0.6\linewidth]{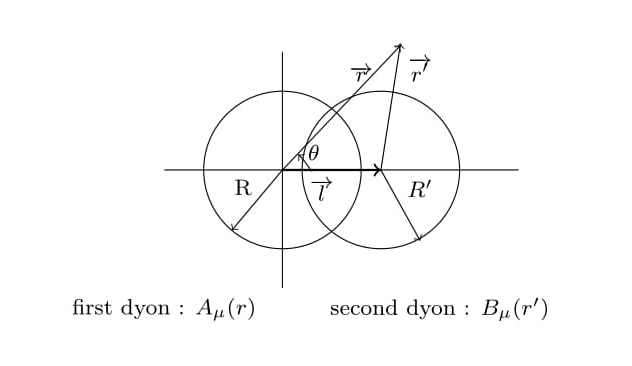}
    \caption{Two overlapping dyons. The first dyon locates at the origin and the second at a distance $l $. $R\sim \frac{1}{\nu}$ and $R' \sim \frac{1}{\alpha}$ are the radius of the dyons where $\nu$ and $\alpha$ are the holonomy of the first and second dyons, respectively. }
       \label{fig:2dyons}
  \end{center}
\end{figure} 

We follow the well-known procedure in optics to study the interaction of optical solitons \cite{Agraw2007}. It is assumed that the solution corresponding to the sum of two solitons is the solution of equation of motion, not the one corresponding to the individual solitons. Using the solution of two solitons system in the self-duality equation (\ref{sd-equation}), some terms are appeared due to the non-linearity of the field strength tensor with respect to the gauge fields. These extra terms are called the "perturbation" terms which are responsible for the interaction between two solitons. These perturbation terms make changes on the original soliton solutions and as a result the shape of the individual dyon deforms. In the following we discuss the details of this phenomena.  

Consider a dyon located at the origin associated with the gauge field $A_{\mu}(r)$ and another dyon at the distance $l \equiv \lvert \overrightarrow{l} \rvert $ with respect to the first dyon defined by the gauge field $ B_{\mu}(r') $, both in Rossi gauge \cite{PRossi82}, as illustrated in figure (\ref{fig:2dyons}). $r'$ denotes the distance from the center of the second dyon located at $ \overrightarrow{l} $. 

As mentioned in the Introduction, we can add two gauge fields of dyons in all gauges, including Rossi gauge, since we do not want to study the dyonic gas. Therefore we do not need to do our calculations in the gauge which minimizes the interaction of dyons.

Assume $ A^{a}_{\mu}(r) + B^{a}_{\mu}(r') $ is the solution of the self-duality equation (\ref{sd-equation}), instead of each individual dyon \cite{Agraw2007}. In this case, the sum ansatz is exactly the solution of Yang-Mills equation and we do not need to be close to the Yang-Mills equation by minimizing the interaction of two dyons. Then we can rewrite this equation with $A^{a}_{\mu}(r)$ and $B^{a}_{\mu}(r')$ functions and their strength tensors $ F^{a}_{\mu \nu} $ and $ G^{a}_{\mu \nu} $, respectively. Using the definition of the strength tensor of equation (\ref{F}) in equation (\ref{sd-equation}) for our two dyon system:
\[ A^{\prime a}_{\mu} =  A^{a}_{\mu} + B^{a}_{\mu}  \]
we have:
\begin{equation}
 \partial _{i} A^{\prime a}_{4} - \partial _{4}A^{\prime a}_{i} + \epsilon _{abc} A^{\prime b}_{i} A^{\prime c}_{4} - \frac{1}{2} \epsilon _{ijk} \left( \partial _{j} A^{\prime a}_{k} - \partial _{k} A^{\prime a}_{j} + \epsilon _{abc} A^{\prime b}_{j} A^{\prime c}_{k} \right) = 0.
\end{equation}
After some calculations, we have
\begin{equation}
\resizebox{\textwidth}{!}{$F^{a}_{i4} - \frac{1}{2}\epsilon _{ijk} F^{a}_{jk} + \epsilon_{abc} A^{b}_{i} B^{c}_{4} - \frac{1}{2} \epsilon _{ijk} \epsilon _{abc} A^{b}_{j}B^{c}_{k} = - \left[ G^{a}_{i4} - \frac{1}{2}\epsilon _{ijk} G^{a}_{jk} + \epsilon_{abc} B^{b}_{i} A^{c}_{4} - \frac{1}{2} \epsilon _{ijk} \epsilon _{abc} B^{b}_{j}A^{c}_{k} \right].$}
\label{fg}
\end{equation}
Since the two sides of equation (\ref{fg}) are equal for all values of $A$ and $B$ functions and for all values of $r$ and $r'$, they should be equal to a constant, which we choose it to be zero \cite{Agraw2007}. Hence we have an equation for the first dyon, $A_{\mu}(r)$
\begin{equation}
F^{a}_{i4} - \frac{1}{2}\epsilon _{ijk} F^{a}_{jk} + \epsilon_{abc} A^{b}_{i} B^{c}_{4} - \frac{1}{2} \epsilon _{ijk} \epsilon _{abc} A^{b}_{j}B^{c}_{k} = 0,
\label{1dyon}
\end{equation}
and the equation for the second dyon is obtained by exchanging $A$ and $B$. Equation (\ref{1dyon}) is the perturbed version of equation (\ref{sd-equation}) for the first dyon by the presence of the second dyon,
\begin{equation}
F^{a}_{i4} - \frac{1}{2}\epsilon _{ijk} F^{a}_{jk} = -\epsilon_{abc} A^{b}_{i} B^{c}_{4} + \frac{1}{2} \epsilon _{ijk} \epsilon _{abc} A^{b}_{j}B^{c}_{k}.
\label{perturbed}
\end{equation}
Therefore, a dyon at the presence of another dyon satisfies equation (\ref{perturbed}) instead of equation (\ref{sd-equation}), with the extra terms on the right hand side of equation (\ref{perturbed}) that we call the perturbation terms. $A_{\mu}(r)$ and $B_{\mu}(r')$ are the gauge fields of individual dyons introduced in equations (\ref{A4}) and (\ref{Ai}).
 
\begin{figure}
 \captionsetup{font=footnotesize}
  \begin{center}
    \includegraphics[width=0.7\linewidth]{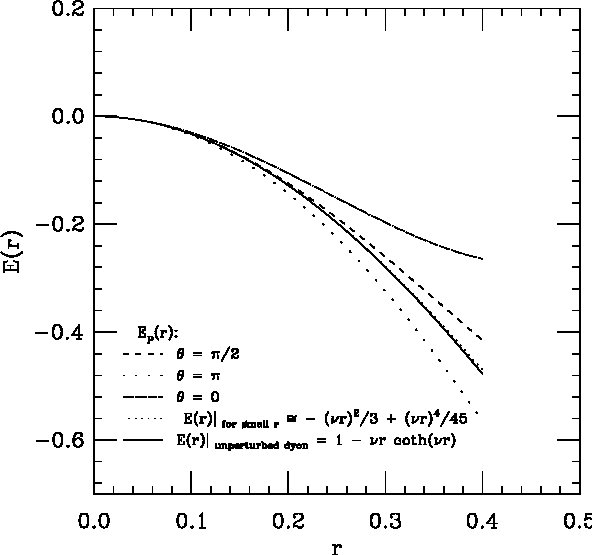}
    \caption{The unperturbed $E(r)$ and perturbed $E_{p}(r)$ for angles $\theta = 0, \pi/2, \pi$ as a function of $r$. Inside the core of the dyon, $ r<\frac{1}{\nu} = \frac{1}{\pi T} \simeq 0.3 \; fm$ and $T$ is fixed to $1 \; fm^{-1}$. For fixed $r$, $E_{p}(r)$ is different for different $\theta$ unlike $E(r)$ which is the same for all $\theta$. We fix the distance between the two dyon centers to $l=0.5 \; fm$ to ensure that the cores of the dyons are overlapped: $0<l<R+R'\simeq 2/\pi$. The plot of the unperturbed $E(r)$ for small $r$ is illustrated to show that the expansion up to $r^{4}$ is a good approximation in the core of the dyon.}
       \label{fig:e}
  \end{center}
\end{figure}

\begin{figure}
 \captionsetup{font=footnotesize}
  \begin{center}
    \includegraphics[width=0.7\linewidth]{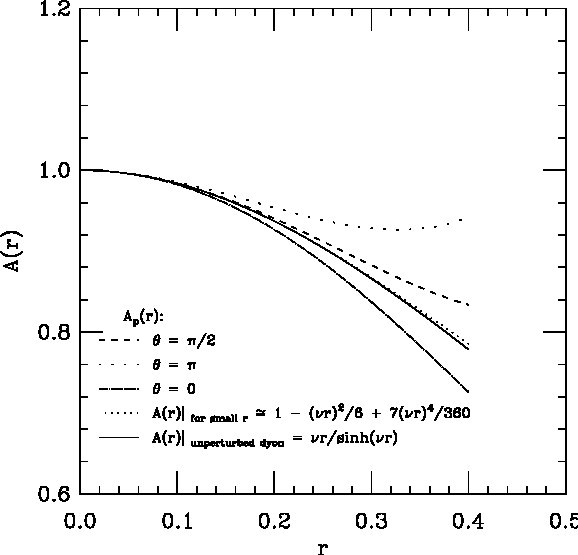}
    \caption{The same as figure (\ref{fig:e}) but for $A(r)$. }
       \label{fig:a}
  \end{center}
\end{figure}

\begin{figure}
\centering
\includegraphics[scale=0.6]{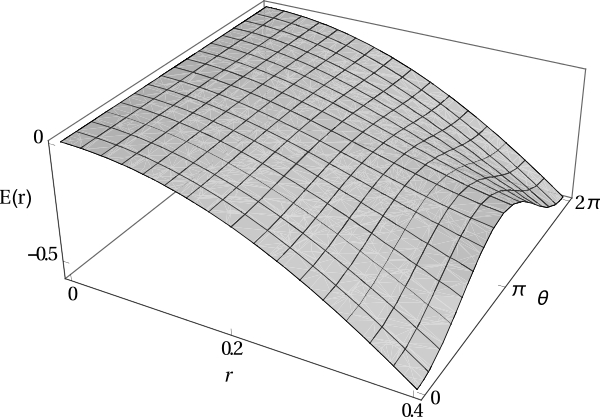}
\includegraphics[scale=0.6]{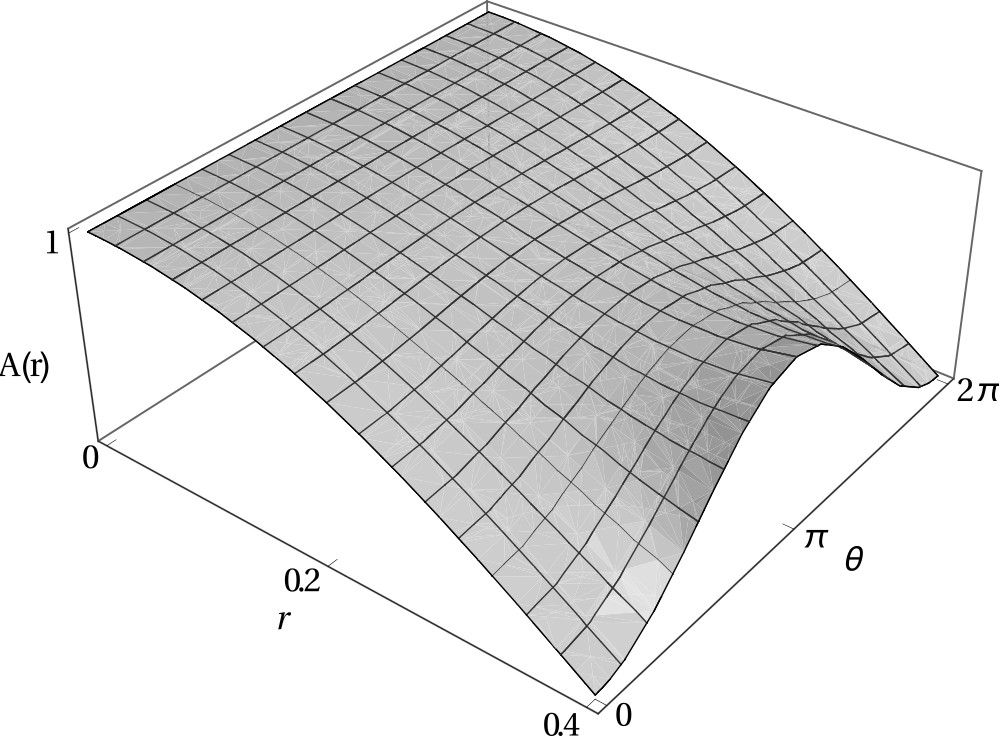}
\caption{$E_{p}(r)$ and $A_{p}(r)$ as a function of $\theta \in [0,2\pi]$. Our approximation is reasonable only in the core of the dyon, $ r<\frac{1}{\nu} = \frac{1}{\pi T} \simeq 0.3 \; fm$ where $T$ is fixed to $1 \; fm^{-1}$. The profile function of the perturbed dyon is a function of $ \theta $ and thus the perturbed dyon has no longer spherical symmetry. }
\label{fig:3ea}
\end{figure}

The left hand side of equation (\ref{perturbed}) can be replaced by equations (\ref{E-field}) and (\ref{B-field}),
\begin{equation}
\begin{split}
LHS & = \left( \delta _{ai} - n_{a}n_{i} \right) \left( \frac{A_{p}(r)}{r}\frac{E_{p}(r)}{r} - \frac{1}{r} \frac{dA_{p}(r)}{dr} \right) \\{}
& + n_{a}n_{i} \left( \frac{d}{dr} \left( \frac{E_{p}(r)}{r} \right) - \frac{A^{2}_{p}(r) - 1}{r^{2}} \right),
\end{split}
\label{lhs}
\end{equation}
where the subscript $p$ denotes the perturbed version of the $A(r)$ and $E(r)$ of equations (\ref{A4}) and (\ref{Ai}). And the perturbation term is responsible for the interaction of these two perturbed dyons. The perturbation terms, right hand side of equation (\ref{perturbed}), can also be rewritten 
\[ \resizebox{1\textwidth}{!}{$ -\epsilon _{abc} A^{b}_{i} B^{c}_{4} = -\epsilon _{abc} \left( \epsilon _{bij} n_{j} \frac{1-A(r)}{r}\right) \left( n_{c} \frac{E(r')}{r'} \right) = \left( \delta _{ai} - n_{a}n_{i} \right) \frac{1-A(r)}{r}\frac{E(r')}{r'}, $}  \]
\[ \resizebox{1\textwidth}{!}{$ \frac{1}{2} \epsilon _{ijk} \epsilon _{abc} A^{b}_{j}B^{c}_{k} = \frac{1}{2} \epsilon _{ijk} \epsilon _{abc} \left( \epsilon _{bjl} n_{l} \frac{1-A(r)}{r}\right) \left( \epsilon _{ckt} n_{t} \frac{1-A(r')}{r'}\right) = n_{a} n_{i} \frac{1-A(r)}{r} \frac{1-A(r')}{r'}, $} \]
\begin{equation}
RHS = \left( \delta _{ai} - n_{a}n_{i} \right) \frac{1-A(r)}{r}\frac{E(r')}{r'} + n_{a} n_{i} \frac{1-A(r)}{r} \frac{1-A(r')}{r'} .
\label{rhs}
\end{equation}
Equating equations (\ref{lhs}) and (\ref{rhs}), we obtain two coupled equations
\begin{equation}
A_{p}(r) \frac{E_{p}(r)}{r} - \frac{d}{dr} A_{p}(r) = \left( 1 - A(r) \right) \frac{E(r')}{r'},
\label{1}
\end{equation}
\begin{equation}
r \frac{d}{dr} \left( \frac{E_{p}(r)}{r} \right) - \frac{A^{2}_{p}(r) - 1}{r} = \left( 1-A(r) \right) \frac{1-A(r')}{r'}.
\label{2}
\end{equation}
As mentioned before, dyons are non-linear and non-Abelian in their cores. Therefore to study equations (\ref{1}) and (\ref{2}), two dyons should be close enough such that their cores overlap. In other words, the distance of their centers in figure (\ref{fig:2dyons}) should be smaller than the sum of their radius, $ l < R + R' $. Hence, we expand $\frac{E(r')}{r'}$ and $\frac{1-A(r')}{r'}$ around $l=0$ to ensure that the cores of two dyons are overlapped. This expansion helps us to rewrite equations (\ref{1}) and (\ref{2}) as a function of $r$.
\begin{equation}
\frac{E(r')}{r'} \simeq \frac{E(r')}{r'}|_{l=0} + l \frac{d}{dl} \left( \frac{E(r')}{r'} \right)|_{l=0}.
\end{equation}
We should find the derivative with respect to $l$,
\[ \frac{d}{dl} \left( \frac{E(r')}{r'} \right) = \frac{dr'}{dl} \frac{d}{dr'} \left( \frac{E(r')}{r'} \right),\]
where the derivative of $r'$ with respect to $l$ is obtained as the following,
\begin{equation}
r' = \lvert \overrightarrow{r} - \overrightarrow{l} \rvert \rightarrow  r' = \left( r^{2} + l^{2} - 2rl \cos \theta \right)^{1/2} \rightarrow \frac{dr'}{dl} = \frac{l - r\cos \theta}{r'},
\end{equation}
$\theta$ is the angle between $ \overrightarrow{r} $ and $ \overrightarrow{l} $ shown in figure (\ref{fig:2dyons}). 
\begin{equation}
\frac{d}{dr'} \left( \frac{E(r')}{r'} \right) = \frac{d}{dr'} \left( \frac{1 - \alpha r' \coth \alpha r'}{r'} \right) = -\frac{1}{r^{\prime2}} + \frac{\alpha^{2}}{\sinh^{2} \alpha r'},
\end{equation}
where we change the parameter $\nu$ in equations (\ref{A4}) and (\ref{Ai}) to $\alpha$ so that the second dyon is distinguished from the first dyon.
\begin{equation}
\frac{d}{dl} \left( \frac{E(r')}{r'} \right)|_{l=0} = \frac{l - r\cos \theta}{r'} \left( -\frac{1}{r^{\prime2}} + \frac{\alpha^{2}}{\sinh^{2} \alpha r'} \right) |_{l=0} = \cos \theta \left( \frac{1}{r^{2}} - \frac{\alpha^{2}}{\sinh^{2} \alpha r} \right),
\end{equation}
therefore,
\begin{equation}
\frac{E(r')}{r'} \simeq \frac{E(r)}{r} + \cos \theta \left( \frac{1}{r^{2}} - \frac{\alpha^{2}}{\sinh^{2} \alpha r} \right) l .
\end{equation}
Now the perturbation term of equation (\ref{1}) can be obtained
\begin{equation}
\left( 1 - A(r) \right) \frac{E(r')}{r'} \simeq  \left( 1 - \frac{\nu r}{\sinh \nu r} \right) \left[ \frac{1 - \alpha r \coth \alpha r}{r} + \cos \theta \left( \frac{1}{r^{2}} - \frac{\alpha^{2}}{\sinh^{2} \alpha r} \right) l \right].
\label{p1}
\end{equation}
Repeating the same procedure, one can find the perturbation term of equation (\ref{2}),
\begin{equation}
\frac{1-A(r')}{r'} \simeq \frac{1-A(r)}{r} + \cos \theta \left( \frac{1}{r^{2}} - \alpha^{2} \frac{\coth \alpha r}{\sinh \alpha r} \right)l,
\end{equation}
\begin{equation}
\left( 1-A(r) \right) \frac{1-A(r')}{r'} \simeq \left( 1 - \frac{\nu r}{\sinh \nu r} \right) \left[ \frac{1}{r} - \frac{\alpha}{\sinh \alpha r} + l\cos \theta \left( \frac{1}{r^{2}} - \alpha^{2} \frac{\coth \alpha r}{\sinh \alpha r} \right) \right].
\label{p2}
\end{equation}
Since we want to find the perturbed version of a dyon in its non-linear region, its core, we should expand the perturbation terms of equations (\ref{p1}) and (\ref{p2}) around $r=0$. Using the expansions of $\coth x$ and $\frac{1}{\sinh x}$ around $x=0$,
\begin{equation}
\coth x \simeq \frac{1}{x} + \frac{x}{3} - \frac{x^{3}}{45} + O(x^{5}),
\label{coth}
\end{equation}
\begin{equation}
\frac{1}{\sinh x} \simeq \frac{1}{x} - \frac{x}{6} + \frac{7x^{3}}{360} - O(x^{5}),
\label{csch}
\end{equation}
we can find the expansion of equations (\ref{p1}) and (\ref{p2}) around $r=0$ up to $r^{4}$,
\begin{equation}
\begin{split}
&\left( 1 - A(r) \right) \frac{E(r')}{r'}  \\{}
&\simeq \left( \frac{\nu^{2}r^{2}}{6} - \frac{7 \nu^{4} r^{4}}{360} \right) \left[ \left( -\frac{\alpha^{2}r}{3} + \frac{\alpha^{4}r^{3}}{45}\right) + l\cos\theta \left( \frac{\alpha^{2}}{3} - \frac{ \alpha^{4} r^{2}}{15} \right) + O(r^{4}) \right] \\{}
& \simeq \frac{\nu^{2}r^{2}}{6} \left[ -\frac{\alpha^{2}r}{3} + l\cos\theta \frac{\alpha^{2}}{3} \right] + O(r^{4}),
\end{split}
\label{pr1}
\end{equation}
\begin{equation}
\begin{split}
&\left( 1-A(r) \right) \frac{1-A(r')}{r'}  \\{}
&\simeq \left( \frac{\nu^{2}r^{2}}{6} - \frac{7 \nu^{4} r^{4}}{360} \right) \left[ \left( \frac{\alpha^{2}r}{6} - \frac{7\alpha^{4}r^{3}}{360}\right) - l\cos \theta \left( \frac{\alpha^{2}}{6} - \frac{21}{360} \alpha^{4} r^{2} \right) + O(r^{4}) \right]\\{}
& \simeq \frac{\nu^{2}r^{2}}{6} \left[ \frac{\alpha^{2}r}{6} - l\cos \theta \frac{\alpha^{2}}{6} \right] + O(r^{4}).
\end{split}
\label{pr2}
\end{equation}
To simplify the notations, let us rename the perturbation terms. We call the right hand sides, of equations (\ref{pr1}) and (\ref{pr2}) $B$ and $C$, respectively,
\begin{equation}
A\frac{E}{r} - A' = B,
\label{B1}
\end{equation}
\begin{equation}
r\frac{d}{dr} \left( \frac{E}{r} \right) - \frac{A^{2}-1}{r} = C,
\label{C1}
\end{equation}
where the functionality and subscript $p$ of $A$ and $E$ are omitted for simplicity. The superscript prime denotes the derivative with respect to $r$. We can find $\frac{E}{r}$ from equation (\ref{B1}),
\[ \frac{E}{r} = \frac{B + A'}{A} \rightarrow r\frac{d}{dr} \left( \frac{E}{r} \right) = \frac{B'}{A} -\frac{BA'}{A^{2}} + \frac{A'' }{A} - \frac{A^{\prime 2}}{A^{2}}, \]
and then putting it in equation (\ref{C2}),
\begin{equation}
r^{2}AB' - r^{2}BA' + r^{2}AA'' - r^{2}A'^{2} - A^{4} + A^{2} - rCA^{2} = 0.
\label{C2}
\end{equation}
Now we should solve the differential equation (\ref{C2}) to find $A$ and put it in equation (\ref{B1}) to find $E$. Solving equation (\ref{C2}) analytically seems difficult if it is possible at all. Hence, we use the polynomial expansion of function $A$, since we are interested in the core region of the dyon, 
\begin{equation}
A = a_{0} + a_{1}r + a_{2}r^{2} + a_{3}r^{3} + a_{4}r^{4} + ...
\label{poly-A}
\end{equation}
Although the expansion (\ref{poly-A}) introduces the perturbed dyon, but it should satisfy the boundary condition $\lim _{r\rightarrow 0} A(r) = 1$, therefore $a_{0} = 1$. Putting expansion (\ref{poly-A}) in equation (\ref{C2}) and after a lengthy but straightforward calculation, one can find the coefficients of expansion (\ref{poly-A}) by equating the coefficient of each power of $r$ with zero. The coefficient $a_{2}$ is unspecified after solving the equation, and we have the freedom to choose it from the unperturbed dyon in (\ref{bc0-A}), therefore we choose $a_{2} = -\frac{\nu^{2}}{6}$.
\[ a_{0} = 1, \;\; a_{1} = 0, \;\; a_{2} = -\frac{\nu^{2}}{6}, \;\; a_{3} = -\frac{5}{144} \nu^{2} \alpha^{2} l \cos \theta, \;\; a_{4} = \frac{7\nu^{2} \left( \alpha^{2} + \nu^{2} \right) }{360},\]
\begin{equation}
A(r) = 1 - \frac{\nu^{2}}{6} r^{2} - \left( \frac{5}{144} \nu^{2} \alpha^{2} l \cos \theta \right) r^{3} + \frac{7\nu^{2} \left( \alpha^{2} + \nu^{2} \right) }{360} r^{4}.
\label{final-A}
\end{equation}
To solve equation (\ref{B1}) to find $E$, we use the polynomial expansion of $E$. Then we use this polynomial and $A$ of equation (\ref{final-A}) in equation (\ref{B1}) to find the coefficients of the polynomial of $E$,
\begin{equation}
E = e_{0} + e_{1}r + e_{2}r^{2} + e_{3}r^{3} + e_{4}r^{4} + ...
\label{poly-E}
\end{equation}
As mentioned in calculating $A$, function $E$ should satisfy the boundary condition $\lim _{r\rightarrow 0} E(r) = 0$, hence $e_{0} = 0$.
\[ a_{0} = 0, \;\; e_{1} = 0, \;\; e_{2} = -\frac{\nu^{2}}{3}, \;\; e_{3} = -\frac{7}{144} \nu^{2} \alpha^{2} l \cos \theta, \;\; e_{4} = \frac{\nu^{2} \left( \alpha^{2} + \nu^{2} \right) }{45},\]
\begin{equation}
E(r) = - \frac{\nu^{2}}{3} r^{2} - \left( \frac{7}{144} \nu^{2} \alpha^{2} l \cos \theta \right) r^{3} + \frac{\nu^{2} \left( \alpha^{2} + \nu^{2} \right) }{45} r^{4}.
\label{final-E}
\end{equation}
Now we can compare the functions of an individual dyon and a dyon at the presence of another dyon in the region of its core,
\begin{subequations}
\begin{align}
E(r) &\simeq -\frac{\nu^{2}}{3} r^{2} + \frac{\nu^{4}}{45} r^{4} \\
E_{p}(r) &\simeq - \frac{\nu^{2}}{3} r^{2} - \left( \frac{7}{144} \nu^{2} \alpha^{2} l \cos \theta \right) r^{3} + \frac{\nu^{2} \left( \alpha^{2} + \nu^{2} \right) }{45} r^{4},
\label{Ep}
\end{align}
\end{subequations}
\begin{subequations}
\begin{align}
A(r) &\simeq 1 -\frac{\nu^{2}}{6} r^{2} + \frac{7 \nu^{4}}{360} r^{4} \\
A_{p}(r) &\simeq 1 - \frac{\nu^{2}}{6} r^{2} - \left( \frac{5}{144} \nu^{2} \alpha^{2} l \cos \theta \right) r^{3} + \frac{7\nu^{2} \left( \alpha^{2} + \nu^{2} \right) }{360} r^{4}.
\label{Ap}
\end{align}
\end{subequations}
To compare the perturbed and unperturbed dyons graphically, we consider two identical dyons such that $ \alpha = \nu $ where $\nu = \pi T$ in the confinement phase and $T$ is the temperature which is fixed to $1$$fm^{-1}$. Figures (\ref{fig:e}) and (\ref{fig:a}) illustrate $E(r)$ and $A(r)$ of an unperturbed dyon for all $r$ and in the limit of $r\rightarrow 0$ and a perturbed dyon for angles $\theta = 0, \pi/2, \pi$. The core size of a dyon $M$ $ \simeq \frac{1}{\nu} $, at confinement phase is of order $\frac{1}{\pi T}$. Since we fix the temperature to $1$$fm^{-1}$, the distance of the centers of two dyon, $0<l<R+R'$, should be smaller than $\frac{2}{\pi} \simeq 0.64$. Therefore we fix $l = 0.5$$fm$ to ensure the cores of the two dyons overlap. These figures show that the expansion of the functions $E(r)$ and $A(r)$ up to $r^{4}$ is good enough, because for $r$ inside the core of the dyon, $r<\frac{1}{\nu} \simeq \frac{1}{\pi} \simeq 0.3  \; fm$, these functions and their expansions are approximately equal. Figure (\ref{fig:3ea}) illustrates the 3D graphs of $E_{p}(r)$ and $A_{p}(r)$ for $\theta\in [0,2\pi]$.

For any given $l$ and fixed $r$ in the core of the dyon, $E_{p}(r)$ and $A_{p}(r)$ in equations (\ref{Ep}) and (\ref{Ap}) change by $\theta$. As a result, the spherical symmetry of the dyons is lost when they overlap each other.
%------------------------------------------------------------------------------------------------------------------------------------------------------------------

\vspace{8ex}

\section{Conclusion}
\label{sec:concl}

We study the interaction of two dyons in the region of their cores. Since dyons are non-linear and non-Abelian objects in their cores, investigating this problem with exact methods is very difficult and complicated. We suppose the presence of a dyon as a perturbation that affects the profile function of another dyon. We use some approximations to ensure that our calculations are in the region of the core of the dyons and also we make sure that the cores of two dyons overlap. Therefore, our calculations are reasonable just in the core. The gauge fields of the perturbed dyon is obtained as a function of the polar angle between radius vector and the vector which connects the centers of two dyons. 

As mentioned in the introduction, our ultimate main goal is to calculate the interaction of the calorons to study the potential of the quark-antiquark pair by interacting caloron ensembles. However, since the interaction of calorons is complex, we have tried to study the interaction of the dyons which are the constituents of the calorons. In the limit where the constituents of the caloron are far away from  each other such that they can be distinguished as two dyons, we can consider the interaction of the dyons as the interaction of the calorons. In that limit, we can make two calorons close enough such that their dyons overlap and study the interaction of the dyons. Of course, the problem is not that easy since the Dirac string between the dyons inside the caloron should be taken care of very carefully and thus more projects and studies are needed to reach the final goal. 

\section{Acknowledgement}
 
We are grateful to the research council of the University of Tehran for supporting this study.
%------------------------------------------------------------------------------------------------------------------------------------------------------------------

\end{document}